# Temporal and Periodic Variation of the MCMESI for the Last Two Solar Cycles; Comparison with the Number of Different Class X-Ray Solar Flares

A. Kilcik[1], P. Chowdhury[2], V. Sarp[1], V. Yurchyshyn[3], B. Donmez[1], J.P. Rozelot[4] A. Ozguc[5]

*(1) Department of Space Science and Technologies, Akdeniz University Faculty of Science, 07058 Antalya, Turkey*

*(2) University College of Science & Technology, Chemical Technology Dept.; University of Cal- cutta; 92, Acharya Prafulla Chandra Road, Kolkata - 700 009, India*

*(3) Big Bear Solar Observatory, Big Bear City, CA 92314, USA*

*(4) Universite Cote d'Azur (UCA), 06130, Grasse, France*

(5) *Kandilli Observatory and Earthquake Research Institute, Bogazici University, 34684 Istanbul, Turkey*

## Abstract

In this study we compared the temporal and periodic variations of the Maximum CME Speed Index (MCMESI) and the number of different class (C, M, and X) solar X-Ray flares for the last two solar cycles (Cycle 23 and 24). To obtain the correlation between the MCMESI and solar flare numbers the cross correlation analysis was applied to monthly data sets. Also to investigate the periodic behavior of all data sets the Multi Taper Method (MTM) and the Morlet wavelet analysis method were performed with daily data from 2009 to 2018. To evaluate our wavelet analysis Cross Wavelet Transform (XWT) and Wavelet Transform Coherence (WTC) methods were performed. Causal relationships between datasets were further examined by Convergence Cross Mapping (CCM) method. In results of our analysis we found followings; 1) The C class X-Ray flare numbers increased about 16 % during the solar cycle 24 compared to cycle 23, while all other data sets decreased; the MCMESI decreased about 16 % and the number of M and X class flares decreased about 32 %. 2) All the X-Ray solar flare classes show remarkable positive correlation with the MCMESI. While the correlation between the MCMESI and C class flares comes from the general solar cycle trend, it mainly results from the fluctuations in the data in case of the X class flares. 3) In general, all class flare numbers and the MCMESI show similar periodic behavior. 4) The 546 days periodicity detected in the MCMESI may not be of solar origin or at least the solar flares are not the source of this periodicity. 5) C and M Class solar flares have a stronger causative effect on the MCMESI compared to X class solar flares. However the only bidirectional causal relationship is obtained between the MCMESI and C class flare numbers.

## 1 Introduction

Coronal Mass Ejections (CMEs) are eruptions of plasma and magnetic field from the solar surface through to the heliosphere. In the course of their travelling in the interplanetary medium at speeds up to 3300 $kms^{-1}$ CMEs interact with the Solar Wind (SW) streams to drive disturbances that have the potential to cause geomagnetic storms (Gosling et al., 1990). Although the possibility and intensity of a geomagnetic storm depends on various parameters such as geometric and kinematic evolution of the incoming disturbance, CMEs are known to be the most significant drivers of space weather in the near-earth space and through the heliosphere.

The formation of a CME typically involves a perturbation in the low corona which is usually in the form of a solar flare or other possible sources such as filament/prominence eruptions, coronal waves or jets (Hudson and Cliver, 2001). Although there is no one-to-one relationship between CME occurrence and solar flares, a causal link between these two eruptive events may be pronounced such that, any flare accompanying a CME is part of an underlying magnetic process (Webb and Howard, 2012). Hence, flares and CMEs can be considered as two different manifestations of the energy release stemming from the rearrangement of the magnetic field in the solar atmosphere. If the released energy is in the form of radiative energy, the resulting phenomena are called a solar flare and if it is in the form of kinetic energy, the resulting phenomena is a CME (Slemzin et al., 2019). But it should be noted that there are secondary conversions between different energy types during solar eruptive processes (Chen, 2011).

CMEs are observed and traced by coronagraphs as a projection of bright expanding structures onto the image plane of the sky. In general, CMEs represent three-part structure, which are bright core in the center of an ejecta, surrounded by a dark cavity with a bright frontal shell facing the interplanetary medium (Cremades and Bothmer, 2004). The outward motion of CMEs in the low plasma-$\beta$ medium is thought to be regulated simultaneously by gas pressure and magnetic forces of solar corona, whose different force-balance interpretations lead to a variety of CME models (Chen, 2011; Byrne et al., 2012). The observational features of CMEs which are used to verify the above mentioned CME models, are mainly morphology, mass, angular width, velocity, acceleration, occurrence rate and energy. Interestingly, kinetic and potential energy of CMEs was found to be comparable to that of solar flares (Emslie et al., 2004).

CME speeds are considered to be a more prominent parameter among others in terms of the effects on space weather. If the speed of a CME with respect to the ambient SW exceeds the local Alfven speed of the corona or interplanetary medium, a shock wave and consequent type II radio burst arise (Cliver et al., 1999). The plasma environment in the interplanetary space is significantly modified by the presence of the CMEs. In order to take into account the disturbances of the background medium (Pomoell and Poedts, 2018) developed a space weather model which uses parameters of the most significant CMEs that observed five days prior to the model run, of the inner heliosphere.

In this paper, we analyzed the temporal and periodic variations of CMEs for the last two Solar Cycles using their maximum speeds and compared the findings with different class X-ray Solar Flares. The outline of the paper is as follows; the specific data to represent CMEs and X-Ray solar flares as well as the methods used during our analyses are discussed in Section 2. The results are given in Section 3. Discussions and Conclusions are given in Section 4.

## 2 Data and Methods

We used CME linear velocity data listed in the Solar and Heliospheric Observatory (SOHO) mission’s Large Angle and Spectrometric Coronagraph (LASCO) CME catalog[1] (Yashiro et al., 2004; Gopalswamy et al., 2009). The catalog covers the period from 1996 until the present. There were total four months gaps in the CME catalog covering 1998 July-September (three months) and 1999 January (one month).

The maximum CME speed index (MCMESI) was first introduced by (Kilcik et al., 2011) as the eruption with the maximum linear fit speed for each day. In the data set the number of observed CME changing from zero to 19 for each day. To obtain the daily MCMESI data, we have chosen a CME that has a maximum linear fit speed from all the CMEs at a given day. Thus we produced the daily MCMESI data. Then we calculated the monthly mean of MCMESI data by taking the average of the daily values for each month. A total of 8401 daily measurements were selected out of a 29469 CME linear fit speed measurements. The monthly version of the MCMESI data were used for cross-correlation, and the daily version of the data were used for periodicity analysis. Selected data that comprise the MCMESI used for the correlation and periodicity analysis. Solar X-ray flare data were taken from the National Oceanographic and Atmospheric Administration (NOAA) Space Weather Prediction Center (SWPC)[2] for the time period from August 1996 through the end of December 2018. Then, the number of C, M, and X class flares were calculated for each day for the investigated time period. For the correlation and periodicity analysis, monthly averaged and the daily values of the above mentioned data sets were used, respectively. Note that due to the four month gap in the CME data, the periodicity analysis utilized data starting February 5, 1999 as suggested by Lou et al (2003).

To investigate the relationship between the maximum CME speeds and different class X-Ray solar flare number data sets used here, we applied the cross-correlation analysis, which determines the correlation coefficient, r, between two time series with possible time delays. To investigate the periodicities of each flare class (C, M, X) and the MCMESI we used Multi Taper Method (MTM) and Morlet wavelet analysis methods.

[1] https://cdaw.gsfc.nasa.gov/CME list/
[2] https://www.swpc.noaa.gov/

The MTM is used to detect low-amplitude harmonic oscillations which have a high degree of statistical significance especially in a given short, noisy as well as multivariate time series. The method uses tapers or orthogonal windows to obtain an estimate of the power spectrum (Thomson, 1982; Ghil et al., 2002). With this method one can also reject larger amplitude harmonics if the F-test fails. This is a main advantage of the MTM over other classical methods where the error bars scale with the peak amplitudes. It has been successfully used for the analysis of climatic data (Ghil et al., 2002; Kilcik et al., 2010; Marullo et al., 2011; Fang et al., 2012; Escudier et al., 2013) and solar data (Kilcik et al., 2018; Chowdhury et al., 2019).

Morlet wavelet analysis, $\psi_n(\eta) = \pi^{-1/4} e^{i\omega_0\eta} e^{-\eta^2/2}$ became a valuable method to analyze localized variations of power within a given time series (Lau and Weng, 1995; Torrence and Compo, 1998). It is a good tool for identifying oscillatory components of a given signal or time series. In contrast to classical Fourier analysis which separates the input signal to its components in the frequency domain, the wavelet analysis transforms 1-D time series into 2-D time-frequency image. It has been widely used in astronomical (Chowdhury et al., 2019; Oloketuyi et al., 2019) and geophysical applications (Torrence and Compo, 1998). In this work, MCMESI and C, M and X flare numbers were analyzed using the Interactive Data Language (IDL) package for Morlet wavelet analysis and the scalograms were obtained to find out the presence and evolution of the periodicities. We have used "Morlet" mother function considering a "red noise" background with a non-dimensional frequency $\omega_0$ = 6 which provides best spatial and temporal resolution (Torrence and Compo, 1998). The effect of edges is represented by the cone of influence (COI) plotted with a bold dotted line. On the other hand, the periods detected above the 95 % confidence level inside the COI are shown by thin black contours. 

To investigate the nonlinear relation and common periods between the MCMESI time series and other flare parameters under study, we further employ the cross-wavelet transform (XWT) and wavelet coherence (WTC) methods. The cross wavelet transform (XWT), for two series [$X_n$] and [$Y_n$] is defined as $W_n^{XY} = Wn^X Wn^{Y*}$, where * represents complex conjugation (Grinsted et al., 2004; Chang and Glover, 2010). The corresponding cross wavelet spectrum $|W_n^{XY}|$ indicates the amount of common power between the two time series as a function of time and frequency. In the XWT spectrum, the phase relation is represented by arrows with the following convention: pointing right, in-phase; pointing left, anti-phase; pointing straight up, the second series Y leads by 90°; pointing straight down, the first series X leads by 90°. A phase mixing between X and Y will arise if the arrows are distributed randomly.

On the other hand, wavelet transform coherence (WTC) measures the cross-correlation between two time series as a function of time and frequency, even though the common power is lows (Maraun and Kurths 2004; Grinsted et al, 2004; Chang and Glover, 2010). The WTC of two time series [$X_n$] and [$Y_n$] is defined as

$$R_n^2(s) = \frac{\left|S\left(s^{-1}W_n^{XY}(s)\right)\right|^2}{S(s^{-1}|W_n^X(s)|^2) \cdot S(s^{-1}|W_n^Y(s)|^2)}$$

where S is a smoothing operator. $R_n^2(s)$ ranges from 0 to 1, and may be considered as a localized correlation coefficient in time–frequency space.We have used the Morlet wavelet as a mother function with $\omega_0 = 6$ to explore both the spectrum of XWT and WTC under the red noise approximation. A Monte-Carlo simulation technique was used to determine the 95 % statistical significance level in all the cross-wavelet plots.

Nonlinear dynamics of the MCMESI and different class solar flare numbers are also investigated by analyzing the causal relations between their state spaces which are described as the set of all possible configurations of a system. State space is the mathematical model of a physical system as a set of input, output and state variables. The state space can be described in the form of the Euclidean space in which the variables on the axes are the state variables. Each state of the system is then represented as a vector within that space. The system state can then be described as a point in a high-dimensional space. The axes of this space can be thought of as fundamental state variables; in this study, these variables correspond to number of different class solar flares and the MCMESI. Note that the system state changes through time following a set of deterministic rules. In other words, the behavior of the system is not completely stochastic.

Because time series are sequential observations of the system behavior, information about the rules that govern system behavior (i.e. the system dynamics) is therefore encoded in the data. Takens' Theorem (Takens, 1981) provides a way to recover this information by reconstructing the state space of a discrete time dynamical system as an E dimensional manifold from the time series data. The optimal choice of the value of the embedding dimension (E) can be tested by the forecasting performance of the state space under a smooth mapping. Here, we used the simplex projection algorithm (Sugihara and May, 1990) to predict the future (known) values (observations) of the datasets and compared them with a constant predictor to find the optimal embedding dimension E. The Simplex Projection algorithm was also used for the solar cycle predictions (Kilcik et al., 2009; Sarp et al., 2018).

One of the corollaries to Takens' Theorem is that multiple reconstructions not only map to the original system, but also to each other. Considering each dataset as a different variable of a common dynamical system, reconstructed state spaces are lower dimensional projections of the common attractor. In case of a causal link between any two variables, cross mappings between their embedded state spaces are expected to cover a considerable portion of the future states in the direction of the causality. Hence, increasing the library size (length of the time series) of the embedded state space will result as a convergence towards higher cross map performance. The method to detect the causality in such way are introduced by Sugihara et al., (2012) as Convergent Cross Mapping (CCM).

## 3 Analysis and Results

### 3.1 Temporal Variations

Figure 1 plots the temporal variation of the MCMESI and the number of flares in each X-

Ray class for the investigated Solar Cycles 23 and 24. All data sets used in this study show about an 11 year solar activity cycle with some differences. The number of flares in X-Ray classes does not exactly follow the average MCMESI during Solar Cycle 23. Also the temporal variations of the number of flares in different classes show differences from each other. On the other hand, flare numbers in all X-Ray classes show similar peaks and follow the same trend during the solar cycle 24, but their temporal evolution is far from being similar to the MCMESI. From this figure it is also seen that the number of M and X class flares decreased significantly (about 32 %) during the solar cycle 24 as compared to cycle 23, while the number of C class flares slightly increased (about 16 %) during the same time period. The MCMESI has also slightly decreased (about 16 %) at the same time period. As shown in this plot there is a notable enhancement of the MCMESI during the declining phase of solar cycle 23 and it is co-temporal with the second peak of the X class flare numbers. The small bump located between 2006–2007 in the MCMESI trend also appears co-temporal with the C and X class flare numbers.

To quantify the relationship between the studied data sets, we applied the cross-correlation analysis and found high that the highest correlation was obtained between the C class flares and the MCMESI (0.72 ± 0.06), while the lowest one was found with X class flare number (0.51 ± 0.09) (see Figure 2 left panel). To remove the contribution of general solar cycle trend into the cross-correlation results, we produced the differenced data and applied the cross-correlation again (see Figure 2, right panel). The differenced data was produced by calculating the differences of consecutive data points. It clearly shows that almost all correlation between the flare number of X class and the MCMESI mainly comes from the fluctuations (0.48 of 0.51), while the correlation between flare number of C class and the MCMESI mostly comes from general trend (0.26 of 0.72).

### 3.2 Periodicity Analysis

To reveal periodic variations in data sets used in this study the MTM and Morlet wavelet analysis results are plotted in Figure 3. A color bar is added to each Morlet wavelet plot which represents the amplitude of power after the wavelet transform. Results of XWT and WTC analysis are shown in Figure 4–6. The detected common quasi-periods above the 95 % confidence level in both XWT and WTC plots are indicated by black contours. In XWT plots, colors represent the amplitude of power and in WTC plots, colors represent the power of the coherence ranging from 0 to 1.

From Figure 3 and Table 2 we conclude that in general, the MCMESI and the X-Ray flares show similar periodicities with some differences: 1) The 1024–1170 days periodicity exist in all data sets except X class flare number. 2) While the 546 days periodicity exists in the MCMESI data with 99 % confidence level, it could not be detected in any class flare number with at least 95 % confidence, thus we may argue that 546 days periodicity detected in the MCMESI may not be solar origin or at least the solar flares is not the source of this periodicity. 3) The 356–390 days periodicity exists in all data sets except C class X-Ray flares, thus we may speculate that the source of this quasi periodicity observed in the MCMESI is the strong solar flare activity. 4) The 303–315 days periodicity detected in the MCMESI possibly comes from the weak solar flares (C class flares). 5) The 227–264 days periodicity was not detected in the MCMESI and C class solar flares. 6) About 75 day periodicity shows exactly the same characteristics with 356–390 days periodicity. Thus it might be the harmonic of 356–390 days periodicity. 7) Solar rotation period (26–38 days) exists in all MTM spectrums and wavelet scalograms of the used data sets, but it is not so prominent in the X class flare number due to the rare occurrence of X class flare event.

Figure 4 represents the wavelet coherence between daily MCMESI and C class flares. It is evident that Rieger-type periods and QBOs are prominent in the range of 1.3 to 3 years. Although there exists a small amount of phase mixing, both MCMESI and C class flares were in phase in case of Rieger-group of periods. In the vicinity of 2–3 year periods, C class flares mostly lead up to 2011 and after that both were in phase. In smaller periods (27–64 days), there exists strong phase mixing. The WTC plot indicates that the correlation coefficient is also high in the above mentioned periodic regions.

The wavelet coherence between daily MCMESI and M class flares are shown in Figure 5. Apart from Rieger type oscillations and Quasi-Biennial Oscillations (QBOs), there exist common periods in the range of solar rotation period and 64—100 days. In case of small and Rieger type periods, in most of the cases both parameters were in phase. In case of QBOs, the M class flare was mostly leading up to 2008 and afterward both again came in phase. Strong correlations have been found for small and Rieger type periods as well as QBOs around 3 years in the WTC spectrum.

The XWT power spectrum for MCMESI and X class flares (Figure 6 upper panel) shows the evidence of high common and significant periods which were already detected in the local wavelet spectra for the individual time series. In the case of Rieger and near Rieger periods as well as periods $\sim$ 1.3 years, both parameters were mostly in phase and highly correlated (Figure 6 lower panel). However, in the range of 280–320 day periods, the MCMESI was leading for the time period of 2006 to 2008. WTC spectrum reveals that a part of the period $\sim$3 years is highly correlated between 2012 and 2014 and after that its power diminished and went out of the COI.

## 3.3 Causal Relationships

Simplex Projection (Sugihara and May, 1990) which is a k-nearest neighbor forecasting algorithm, was used to determine the optimal embedding dimensions for state space reconstructions. First, each time series used in this study was divided into two parts such that 2/3 of the data points were used as library points for embedded state space reconstructions and the remaining 1/3 was used as observations to compare with the predicted values and calculate Pearson’s correlation coefficient ($\rho$). Assuming a low dimensional (up to 13 dimension) common attractor, the results of simplex projection algorithm with varying embedding dimension values were compared with a naive constant predictor (where the 1-step ahead forecast is the current value) over the same set of predictions models. Results are given in (Figure 7) where the vertical axes represent the correlation coefficients between the predicted and observed values for both simplex projection (red) and the constant predictor (black). Optimal embedding dimensions for the reconstructions of the embedded state spaces are searched to be as low as possible and have a considerably better prediction performance than constant predictor. Optimal embedding values are thus determined 6 for the MCMESI, Number of C and M Class Solar Flare time series, and 9 for the number of X Class Solar Flares. This higher embedding dimension for the Number of X Class Solar Flare time series compared to the others indicates a more complex state space is needed to unfold the true dynamics of X Class Solar Flares.

As a second step of analyzing the causal relations, state space of the MCMESI is mapped towards the number of different class solar flare time series. Cross mapping scores are calculated as in Simplex Projection mentioned above. The results are averaged and plotted as a function of library size in the left panel of Figure 8. The results show clear evidence of convergence for the MCMESI mapping the different class solar flares, with the maximum cross mapping skill, ($\rho = 0.74$) for the C class solar flares is slightly higher than its cross correlation ($r = 0.72$) with the MCMESI. Same procedure is applied with the inverse mapping direction where numbers of different class solar flares are used as libraries and mapped towards the MCMESI. The results again show the convergence for the cross map scores as can be seen in the right panel of Figure 8. In other words, dynamics of the embedded state space of the MCMESI is sensitive to the variations of number of each class solar flares. However, numbers of M and X Class solar flares are found to have a higher mapping performance when used as the library set instead of the target set. This finding indicates a unidirectional causality or a common driver which is more effective on the numbers of M and X Class solar flares compared to the MCMESI.

## 4 Conclusions and Discussions

In this study we analyzed the temporal and periodic variations of the MCMESI and compared with the different classes X-Ray solar flare numbers. We found the followings;

- The number of C class X-Ray flares increased about 16 % during the solar cycle 24 compared to cycle 23, while all other data sets decreased; the MCMESI decreased about 16 % and the number of M and X class flares decreased about 32 %.
- During solar cycle 23 only C and M flare peaks correspond to each other and they show different behavior from the MCMESI, while during solar cycle 24 the MCMESI and flare peaks coincide in time their amplitudes do not match; for the MCMESI the first peak stronger but for the flares the situation is opposite.
- All X-Ray solar flare classes show remarkable positive correlation with the MCMESI. While the correlation between the MCMESI and C class flares comes from the general solar cycle trend, it mainly comes from the fluctuations in the data in case of the number of X class flares.
- The number of flares in all classes and the MCMESI shows similar periodic behavior during the investigated time period.
- The 546 days periodicity detected in the MCMESI is not found in any class of X-Ray solar flare. Thus, we may argue that this periodicity has no X-Ray solar flare origin or at least the X-Ray solar flares are not the source of this periodicity for the last two solar cycles.
- In general Rieger type and longer periodicities are in phase in both the MCMESI and flare data sets, however the short-term periodicities show more phase mixing.
- C and M Class Solar Flares have a stronger causative effect on the MCMESI compared to X class solar flares. However the only bidirectional causal relation is between the MCMESI and the number of C class flares.

In this paper we have investigated the temporal and periodic variations of the MCMESI and X-Ray solar flare numbers of C, M, and X classes. Although the highest correlation between the MCMESI and X-Ray solar flare numbers are found for the C class flares, this relation was lost when the general trend of solar cycle was removed. On the other hand, the correlation between the MCMESI and X class flares are found to be more meaningful than other classes since it was shown by comparing the differenced data (see Figure 2, right panel). So, the general trend of solar cycle has a negligible impact on the correlation between X class solar flare numbers and the MCMESI. This was also seen in the correlation coefficients of differenced data sets that gave the highest value for the X class flares.

It has been long known that the fastest CMEs display an almost constant speed during their evolution (MacQueen and Fisher, 1983; Andrews and Hovard, 2001). This lack of velocity variation during a fast CME propagation allowed us to use the linear speed as a single event metric in the form of the MCMESI time series. On the other hand CMEs undergo three phases during their kinematic evolutions which are initiation, impulsive acceleration, and propagation phases, respectively (Zhang et al., 2001). The initiation phase, which determines the initial speeds of CMEs, is comprised by a slow ascension period prior to the impulsive phase (Kundu et al., 2004) but X class solar flare related CMEs are shown to lack the slow ascension phase during their initiation (Zhang et al., 2001). This phenomenon may explain the reason for the highest correlation of X class flares in the differenced data in this study. Our results are in agreement with the above mentioned findings such that, the fastest CMEs, which do not undergo the slow ascension phase, are mainly related to the X class solar flares and the general trend of the solar cycle has no effect on the coupled correlation between the MCMESI and X class solar flare numbers.

The results of this study indicate that the periodicities lower than 0.5 years are more common for all analyzed data. QBOs, which are found to be related with the turbulent $\alpha$–dynamo (Inceoglu et al., 2019) are also observed for the MCMESI although some periods in the QBO range are not present in each type of flares. For instance, the X class solar flares have the maximum periodicity of roughly one year while neither class of flares display 546-day periodicity. It is reported that the 540-day periodicity is due to the emergence of magnetic flux in solar cycles (Oliver et al., 1992). But, in our analysis, we could not found 546-days periodicity in any class of X-ray solar flare for the last two solar cycles and it is found only in the MCMESI data. Thus, we may argue that this periodicity has no X-ray solar flare origin or at least the solar flares are not the source of this periodicity for the last two solar cycles. It is also reported that the CME initiation and acceleration are known to be connected with the solar flare mechanism (Zhang et al., 2001), we may conclude that this connection encapsulates the interference of different types of flares with CMEs with various speeds. The exact reason behind the observed mid-term periodicities is still unclear. However, it is possible that Rieger type periods may be linked with the periodic emergence of magnetic flux from deep solar interior which triggers the periodic occurrence of solar flare events (Ballester et al., 2002) or probably related with the dynamics/excitation of unstable, magnetic Rossby types of waves ('r' mode oscillations) inside the tachocline (Zaqarashvili et al., 2010; Gurgenashvili et al., 2016; Zaqarashvili and Gurgenashvili, 2018; Chowdhury et al., 2019).

Recently, different parameters of CMEs such as number and mass are studied for their periodicity and very few common periods are found between the CMEs and other solar activity events (Barlyaeva et al., 2018; Lamy et al., 2019) and mainly the reported common periods are less than one year. Lamy et al., (2019) concluded that the most important common periods between the CME occurrence rates and sunspot areas are around 3.2 months. The 3.2 months periodicity which corresponds to 100-120 day periodicity in our results agrees well with all types of solar flares as well as the MCMESI. Barlyaeva et al., (2018) further reported 2.1, 4.4, 8.9 months periodicities for M- Class flares and 4.6, and 7.5 months periodicities for X-class flares. All these reported periodicities are consistent with our results. These authors also reported 2.4-year ($\sim$ 880 days) periodicity for the CME number and 1.7-year ($\sim$ 620 days) periodicity for the CME mass. In this study, the corresponding results to these values are 1024–1071 days and 546 days, respectively. Thus, it can be said that not all the parameters of CME events show the same range of periodicities. Our Morlet wavelet spectra indicates that except MCMESI, other different class of flares practically have no period during the epoch 2008 to 2010 which was the deep minima of solar cycle 23/24. Along with it, our XWT and WTC plots exhibit the absence of small periods including Rieger type and low correlation in time–frequency space during this epoch. This is due to the fact that during low solar activity the flare signals were very poor because of the low numbers of sunspots and active regions.

Papaioannou et al., (2016) analyzed 888 CME-solar flare pairs that have been observed during the SOHO era (1996–2013) in order to extract empirical relations of Solar Energetic Particle (SEP) events to their parent solar sources. Their sample consists of 585 C, 235 M and 68 X class solar flares. Analyzing this sample, they revealed the nonlinear relation of the enhanced radiation levels dependence on CME velocity and solar flare magnitude (See Figure 21 panel A in their paper). Based on their relative timings these authors further indicated a causal relationship between CMEs and solar flares. However taking into account the fact that correlation does not necessarily imply causation, we extended our analyses by reconstructing state spaces of the studied variables. Thus, our results disagree with the above mentioned findings. Although CCM results show the clear causal relation between the MCMESI and C class flare numbers, the unidirectionality of M and X class flare numbers indicates the studied synthetic index, MCMESI is not significative on the M and X class flare numbers. But it should also be noted that these authors used the mean speed of the CMEs during their analyses while the MCMESI consist of only the maxima.

Another possible explanation to the asymmetries of CCM results may be a common driver with varying effects on different phenomenon. Figure 8 revealed that C and M class flare numbers both have a similar and strong causal link towards the MCMESI. In addition, X class solar flare numbers also have a considerable cross map skill towards the MCMESI compared to the reverse mapping direction. Considering these findings, it is clear that the only asymmetric (or equivalent unidirectional) causal relation is valid for the number of C Class solar flares and the MCMESI. The hint for a possible common driver between two data sets are also revealed at Figure 2 where the high cross correlation coefficients between these two are found due to a common seasonal (cyclic) variation and the differenced data sets display a very poor correlation. CCM results agree with this finding since mapping from/towards the MCMESI does not affect the score for C class solar flares. In order to reveal and analyze the cyclic variation, more indices for solar flares and coronal mass ejections are needed to be cross mapped which is not the main focus for this study.

**References**

Andrews, M. D. and Hovard, R. A. (2001). A two-Type Classification of Lasco Coronal Mass Ejection. *Space Science Reviews*, 95:147–163.

Ballester, J. L., Oliver, R., and Carbonell, M. (2002). The Near 160 Day Periodicity in the Photospheric Magnetic Flux. *Astrophys. J.*, 566:505.

Barlyaeva, T., Wojak, J., Lamy, P., Boclet, B., and Toth, I. (2018). Periodic behaviour of coronal mass ejections, eruptive events, and solar activity proxies during solar cycles 23 and 24. *J. Atmos. Solar-Terr. Phys.*, 177:12–28.

Byrne, J. P., Morgan, H., Habbal, S. R., and Gallagher, P. T. (2012). Auto- matic Detection and Tracking of Coronal Mass Ejections. II. Multiscale Filtering of Coronagraph Images. *Astrophys. J.*, 752(2):145.

Chang, C. and Glover, G. H. (2010). Time-frequency dynamics of resting- state brain connectivity measured with fMRI. *NeuroImage*, 50(1):81–98.

Chen, P. F. (2011). Coronal Mass Ejections: Models and Their Observational Basis. *Living Reviews in Solar Physics*, 8(1):1.

Chowdhury, P., Kilcik, A., Yurchyshyn, V., Obridko, V. N., and Rozelot, J. P. (2019). Analysis of the Hemispheric Sunspot Number Time Series for the Solar Cycles 18 to 24. *Solar Phys.*, 294(10):142.

Cliver, E. W., Ling, A. G., Wise, J. E., and Lanzerotti, L. J. (1999). A predic- tion of geomagnetic activity for solar cycle 23. *J. Geophys. Res.*, 104(A4):6871–6876.

Cremades, H. and Bothmer, V. (2004). On the three-dimensional configuration of coronal mass ejections. *Astron. Astrophys.*, 422:307–322.

Emslie, A. G., Kucharek, H., Dennis, B. R., Gopalswamy, N., Holman, G. D., Share, G. H., Vourlidas, A., Forbes, T. G., Gallagher, P. T., Mason, G. M., Metcalf, T. R., Mewaldt, R. A., Murphy, R. J., Schwartz, R. A., and Zurbuchen, T. H. (2004). Energy Partition in Two Solar Flare/CME Events. *J. Geophys. Res.*, 109(A10):10104.

Escudier, R., Mignot, J., and Swingedouw, D. (2013). A 20-year coupled ocean-sea ice-atmosphere variability mode in the North Atlantic in an AOGCM. *Climate Dy- namics*, 40(3–4):619–636.

Fang, K., Gou, X., Chen, F., Liu, C., Davi, N., Li, J., Zhao, Z., and Li, Y. (2012). Tree-ring based reconstruction of drought variability (1615-2009) in the Kongtong Mountain area, northern China. *Global and Planetary Change*, 80:190–197.

Ghil, M., Allen, M. R., Dettinger, M. D., Ide, K., Kondrashov, D., Mann, M. E., Robertson, A. W., Saunders, A., Tian, Y., Varadi, F., and Yiou, P. (2002). Advanced spectral methods for climatic time series. *Reviews of Geophysics*, 40(1):1003.

Gopalswamy, N., Yashiro, S., Michalek, G., Stenborg, G., Vourlidas, A., Freeland, S., and Howard, R. (2009). The SOHO/LASCO CME catalog. *Earth Moon and Planets*, 104(1–4):295–313.

Gosling, J. T., Bame, S. J., McComas, D. J., and Phillips, J. L. (1990). Coronal mass ejections and large geomagnetic storms. *Geophys. Res. Lett.*, 17(7):901–904.

Grinsted, A., Moore, J. C., and Jevrejeva, S. (2004). Application of Cross Wavelet Transform and Wavelet Coherence to Geophysical Time Series. *Nonlinear Processes in Geophysics*, 11:561–566.

Gurgenashvili, E., Zaqarashvili, T. V., Kukhianidze, V., Oliver, R., Ballester, J. L., Ramishvili, G., Shergelashvili, B., Hanslmeier, A., and Poedts, S. (2016). Rieger- type Periodicity during Solar Cycles 14-24: Estimation of Dynamo Magnetic Field Strength in the Solar Interior. *Astrophys. J.*, 826(1):55.

Hudson, H. S. and Cliver, E. W. (2001). Observing coronal mass ejec- tions without coronagraphs. *J. Geophys. Res.*, 106(A11):25199–25214.

Inceoglu, F., Simoniello, R., Arlt, R., and Rempel, M. (2019). Constraining non-linear dynamo models using quasi-biennial oscillations from sunspot area data. *Astron. Astrophys.*, 625:A117.

Kilcik, A., Anderson, C. N. K., Rozelot, J. P., Ye, H., Sugihara, G., and Ozguc, A. (2009). Nonlinear Prediction of Solar Cycle 24. *Astrophys. J.*, 693(2):1173–1177.

Kilcik, A., Ozguc, A., and Rozelot, J. P. (2010). Latitude dependency of solar flare index–temperature relation occuring over middle and high latitudes of Atlantic–Eurasian region. *J. Atmos. Solar-Terr. Phys.*, 72(18):1379–1386.

Kilcik, A., Yurchyshyn, V., Sahin, S., Sarp, V., Obridko, V., Ozguc, A., and Rozelot, J. P. (2018). The evolution of flaring and non-flaring active regions. *Mon. Not. Roy. Astron. Soc.*, 477(1):293–297.

Kilcik, A., Yurchyshyn, V. B., Abramenko, V., Goode, P. R., Gopalswamy, N., Ozguc, A., and Rozelot, J. P. (2011). Maximum Coronal Mass Ejection Speed as an Indicator of Solar and Geomagnetic Activities. *Astrophys. J.*, 727(1):44.

Kundu, M. R., White, S. M., Garaimov, V. I., Manoharan, P. K., Subrama- nian, P., Ananthakrishnan, S., and Janardhan, P. (2004). Radio Observations of Rapid Ac- celeration in a Slow Filament Eruption/Fast Coronal Mass Ejection Event. *Astrophys. J.*, 607(1):530–539.

Lamy, P. L., Floyd, O., Boclet, B., Wojak, J., Gilardy, H., and Barlyaeva, T. (2019). Coronal Mass Ejections over Solar Cycles 23 and 24. *Space Sci. Rev.*, 215(5):39.

Lau, K. M. and Weng, H. (1995). Climate Signal Detection Using Wavelet Transform: How to Make a Time Series Sing. *Bulletin of the American Meteorological Society*, 76(12):2391–2402.

MacQueen, R. M. and Fisher, R. R. (1983). The kinematics of solar inner coronal transients. *Solar Phys.*, 89(1):89–102.

Maraun, D. and Kurths, J. (2004). Cross wavelet analysis: Significance testing and pitfalls. *Nonlinear Processes in Geophysics*, 11(4):505–514.

Marullo, S., Artale, V., and Santoleri, R. (2011). The SST Multidecadal Variability in the Atlantic–Mediterranean Region and Its Relation to AMO. *Journal of Climate*, 24(16):4385–4401.

Oliver, R., Carbonell, M., and Ballester, J. (1992). Intermediate-term period- icities in solar activity. *Solar Phys.*, 137:141.

Oloketuyi, J., Liu, Y., and Zhao, M. (2019). The Periodic and Temporal Behaviors of Solar X-Ray Flares in Solar Cycles 23 and 24. *Astrophys. J.*, 874(1):20.

Papaioannou, A., Sandberg, I., Anastasiadis, A., Kouloumvakos, A., Georgoulis, M. K., Tziotziou, K., Tsiropoula, G., Jiggens, P., and Hilgers, A. (2016). Solar flares, coronal mass ejections and solar energetic particle event characteristics. *Journal of Space Weather and Space Climate*, 6:A42.

Pomoell, J. and Poedts, S. (2018). EUHFORIA: European heliospheric forecasting information asset. *Journal of Space Weather and Space Climate*, 8:A35.

Sarp, V., Kilcik, A., Yurchyshyn, V., Rozelot, J. P., and Ozguc, A. (2018). Prediction of solar cycle 25: a non-linear approach. *Mon. Not. Roy. Astron. Soc.*, 481(3):2981– 2985.

Slemzin, V. A., Goryaev, F. F., Rodkin, D. G., Shugay, Y. S., and Kuzin, S. V. (2019). Formation of Coronal Mass Ejections in the Solar Corona and Propagation of the Resulting Plasma Streams in the Heliosphere. *Plasma Physics Reports*, 45(10):889–920.

Sugihara, G., May, R., Ye, H., Hsieh, C.-h., Deyle, E., Fogarty, M., and Munch, S. (2012). Detecting Causality in Complex Ecosystems. *Science*, 338(6106):496.

Sugihara, G. and May, R. M. (1990). Nonlinear forecasting as a way of distinguishing chaos from measurement error in time series. *Nature*, 344(6268):734–741.

Takens, F. (1981). Detecting strange attractors in turbulence. *Lecture Notes in Mathematics*, 898:366.

Thomson, D. J. (1982). Spectrum Estimation and Harmonic Analysis. *IEEE Proceedings*, 70:1055–1096.

Torrence, C. and Compo, G. P. (1998). A Practical Guide to Wavelet Analysis. *Bull. Am. Meteorol. Soc.*, 79:61–78.

Webb, D. F. and Howard, T. A. (2012). Coronal Mass Ejections: Ob- servations. *Living Reviews in Solar Physics*, 9(1):3.

Yashiro, S., Gopalswamy, N., Michalek, G., St. Cyr, O. C., Plunkett, S. P., Rich, N. B., and Howard, R. A. (2004). A catalog of white light coronal mass ejections observed by the SOHO spacecraft. *J. Geophys. Res.*, 109:A07105.

Zaqarashvili, T. V., Carbonell, M., Oliver, R., and Ballester, J. L. (2010). Quasi-biennial Oscillations in the Solar Tachocline Caused by Magnetic Rossby Wave Instabili- ties. *Astrophys. J. Lett.*, 724(1):L85–L98.

Zaqarashvili, T. V. and Gurgenashvili, E. (2018). Magneto- Rossby waves and seismology of solar interior. *Frontiers in Astronomy and Space Sciences*, 5:7.

Zhang, J., Dere, K. P., Howard, R. A., Kundu, M. R., and White, S. M. (2001). On the Temporal Relationship between Coronal Mass Ejections and Flares. *Astrophys. J.*, 559(1):452–462.

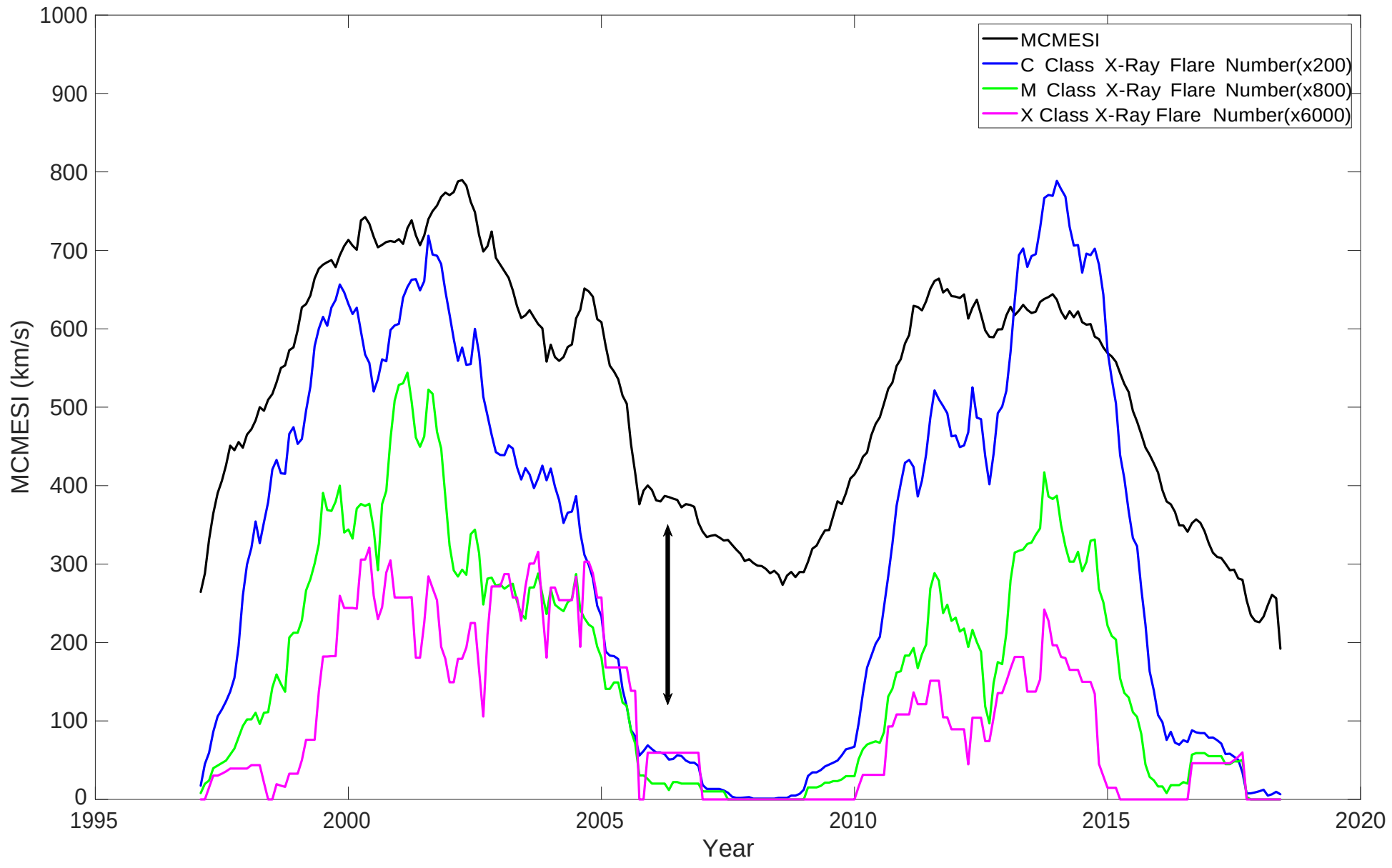

Figure 1: Temporal variations of the MCMESI and different class X-Ray solar flares for the investigated time period.

Table 1: Cross correlation coefficients between each flare class and the MCMESI for monthly and differenced data sets.

| | C | M | X |
|---|---|---|---|
| Monthly Data | 0.72 ± 0.06 | 0.63 ± 0.08 | 0.51 ± 0.09 |
| Differenced Data | 0.26 ± 0.12 | 0.37 ± 0.11 | 0.48 ± 0.1 |

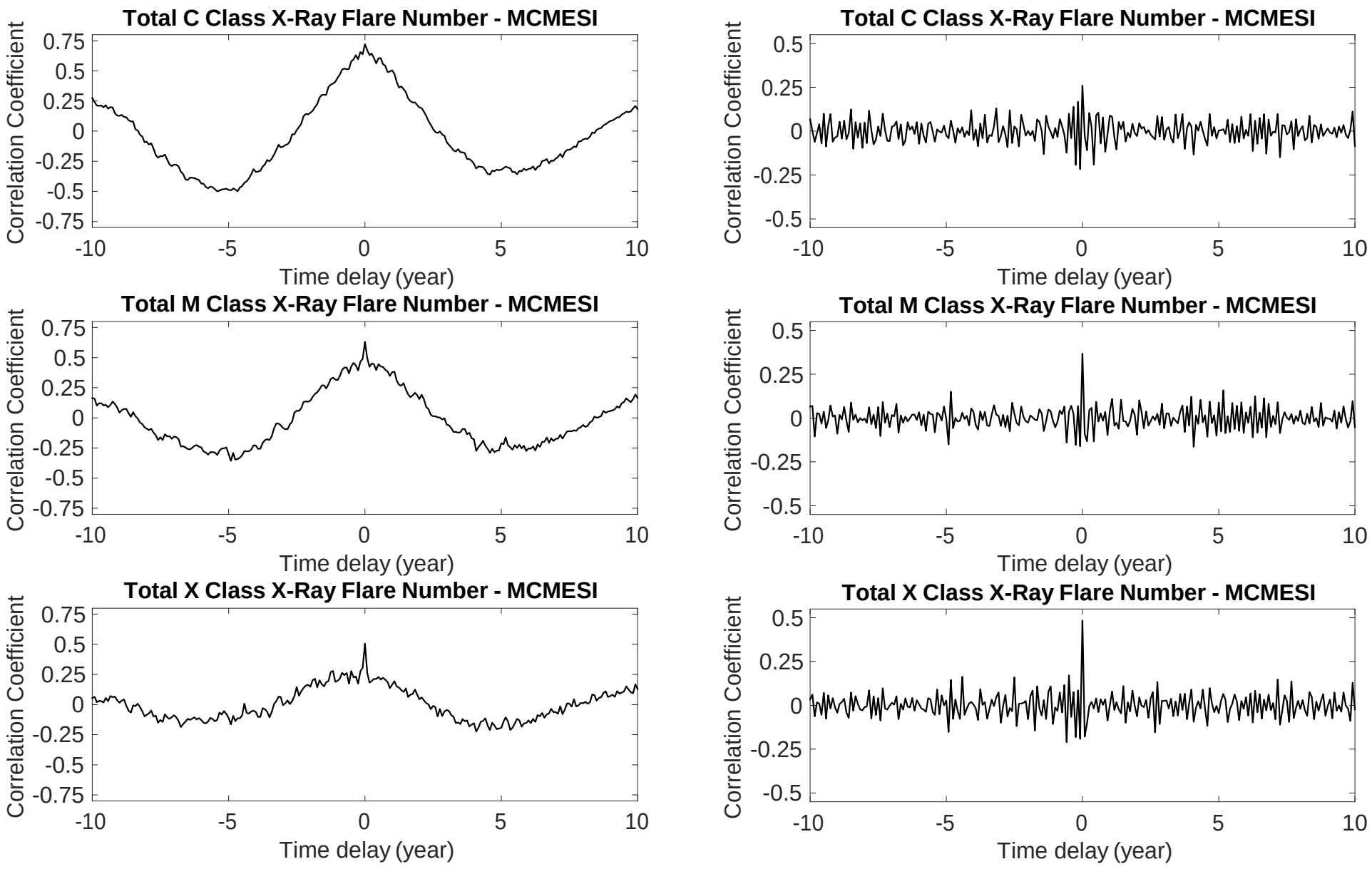

Figure 2: Cross-correlation analysis results between the MCMESI and different class X-Ray flare numbers (left panel) and differenced data sets (right panel).

Table 2: Detected meaningful periods and their existence and significance level in each data set used in this study.

| Period [*Day*] | MCMESI | C Class Flare Number | M Class Flare Number | X Class Flare Number |
|---|---|---|---|---|
| 1024–1170 | + > 99 | + > 99 | + > 99 | – |
| 546 | + > 99 | – | – | – |
| 365–390 | + > 99 | – | + > 95 | + > 95 |
| 303–315 | + > 99 | + > 99 | – | – |
| 227–264 | – | – | + > 99 | + > 99 |
| 134–171 | + > 95 | + > 95 | + > 95 | + > 95 |
| 100–120 | + > 95 | + > 99 | + > 95 | + > 95 |
| 83–95 | + > 99 | + > 95 | + > 95 | + > 95 |
| 68–76 | + > 99 | – | + > 99 | + > 99 |
| 52–62 | + > 99 | + > 99 | + > 95 | + > 95 |
| 39–46 | + > 99 | + > 95 | + > 95 | + > 99 |
| 26–37 | + > 99 | + > 95 | + > 95 | + > 95 |

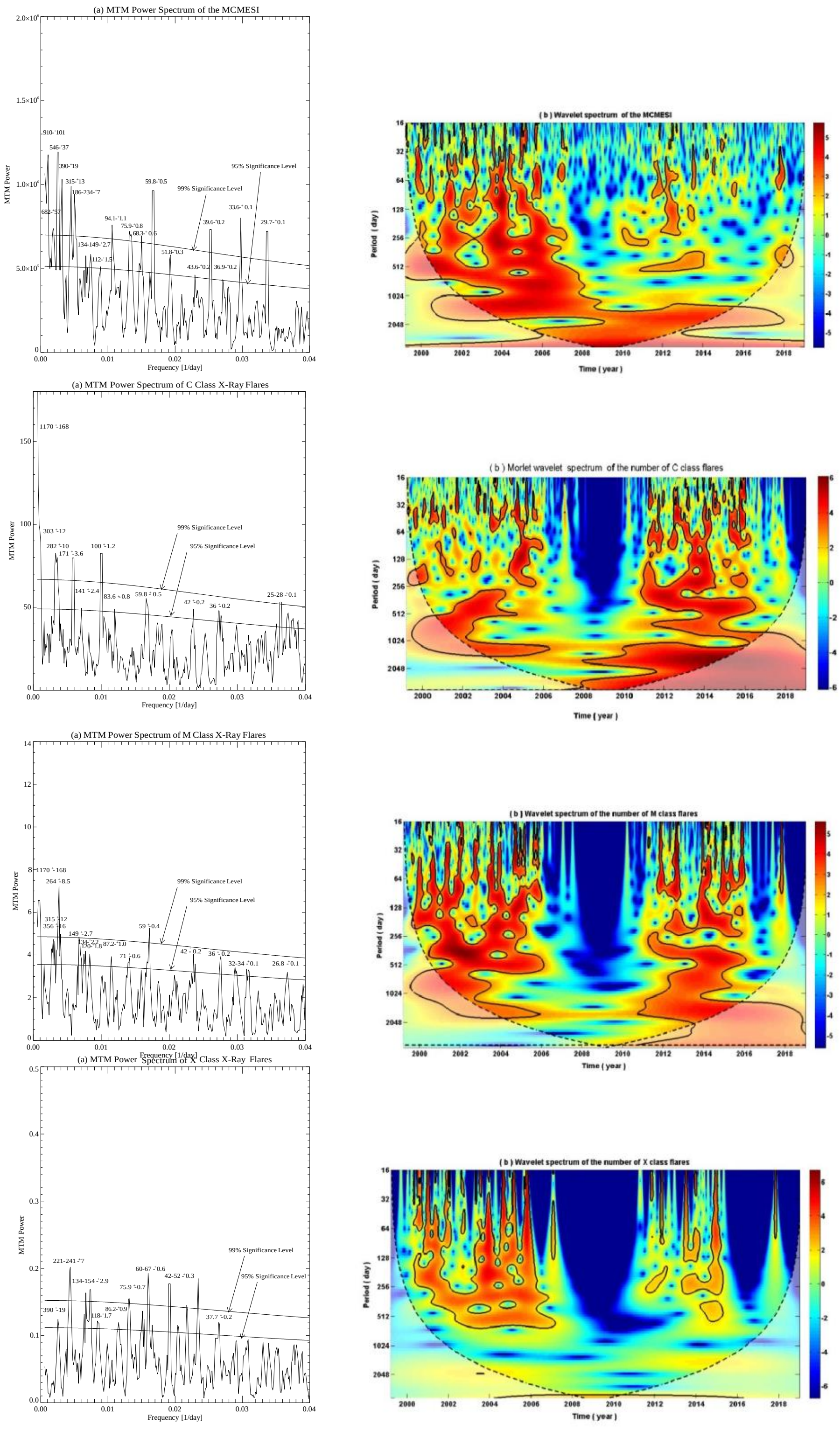

Figure 3: Left panels show the results of multi-taper method of the MCMESI, C, M, and X classes flare numbers respectively from top to bottom. Right panel show the Morlet analysis results of the same data sets, respectively. The numbers in the MTM spectrum peaks indicate the period/errors in days. The 95 % and 99 % confidence levels are shown by solid curves in these plots. The 95 % confidence level in the wavelet scalogram is indicated by the black contours and the cone of influence marked with the area below the dashed line.

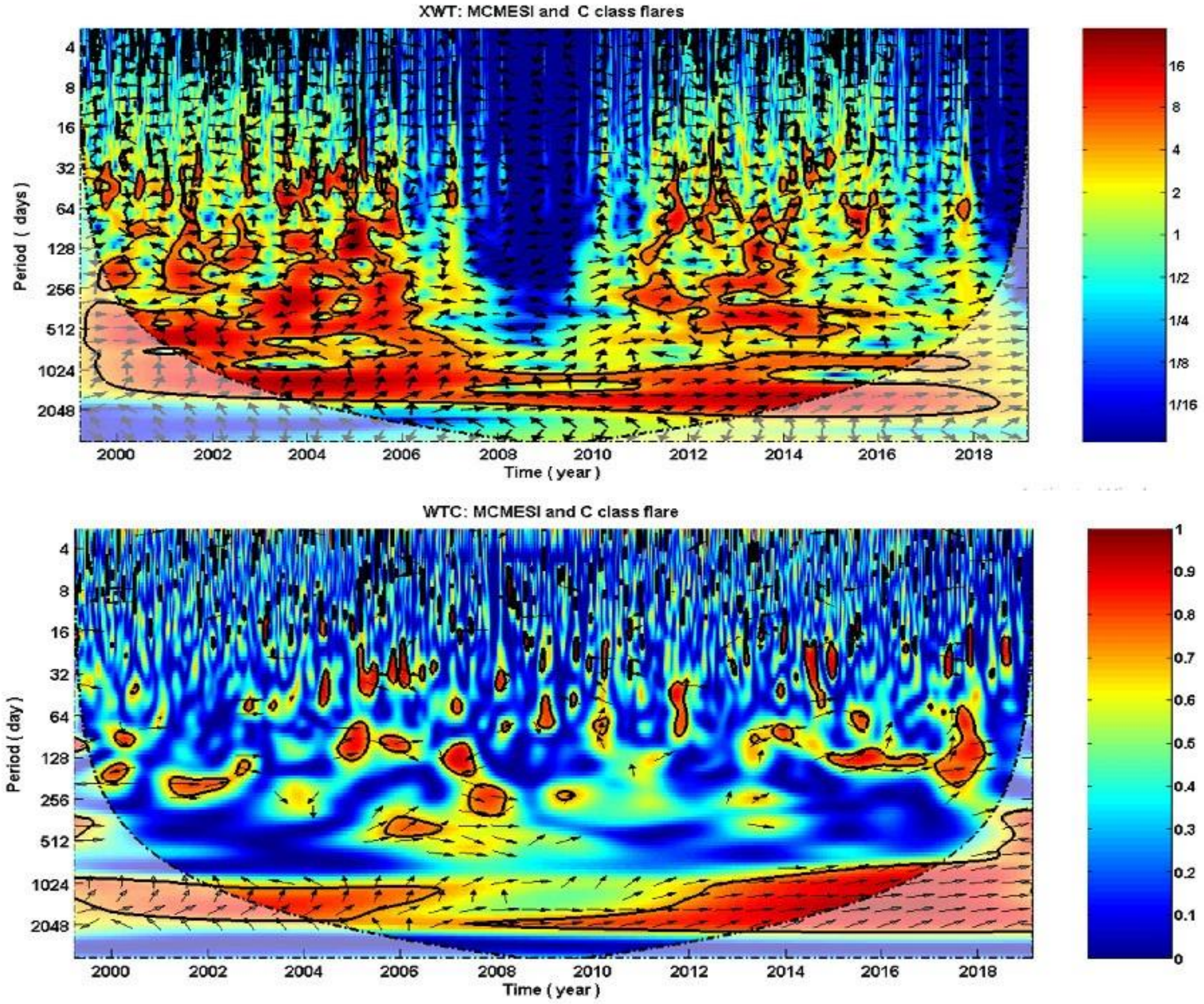

Figure 4: XWT (upper panel) and WTC (lower panel) spectrum between daily MCMESI and C class flares.

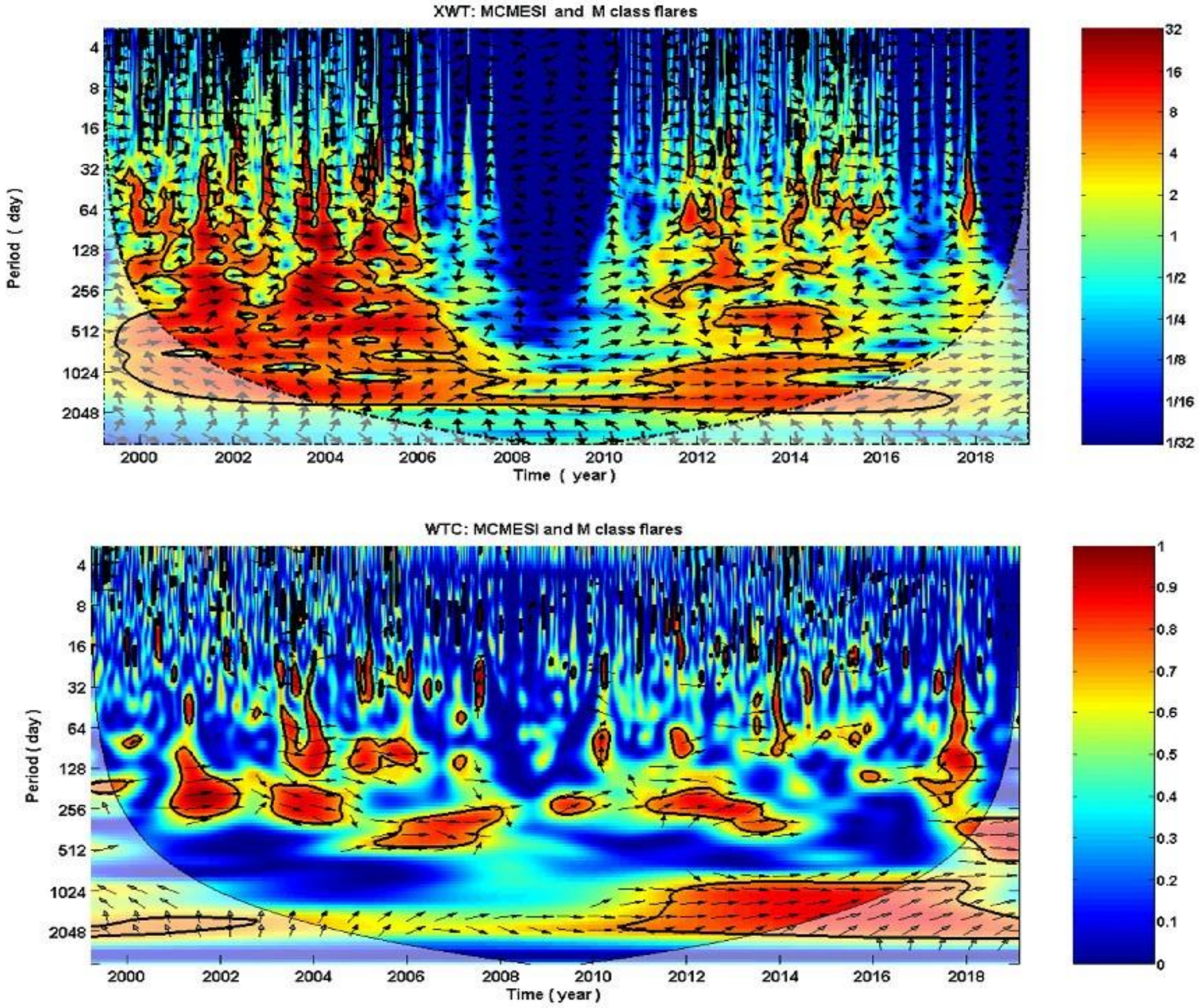

Figure 5: The same as Figure 4 but for M class flares.

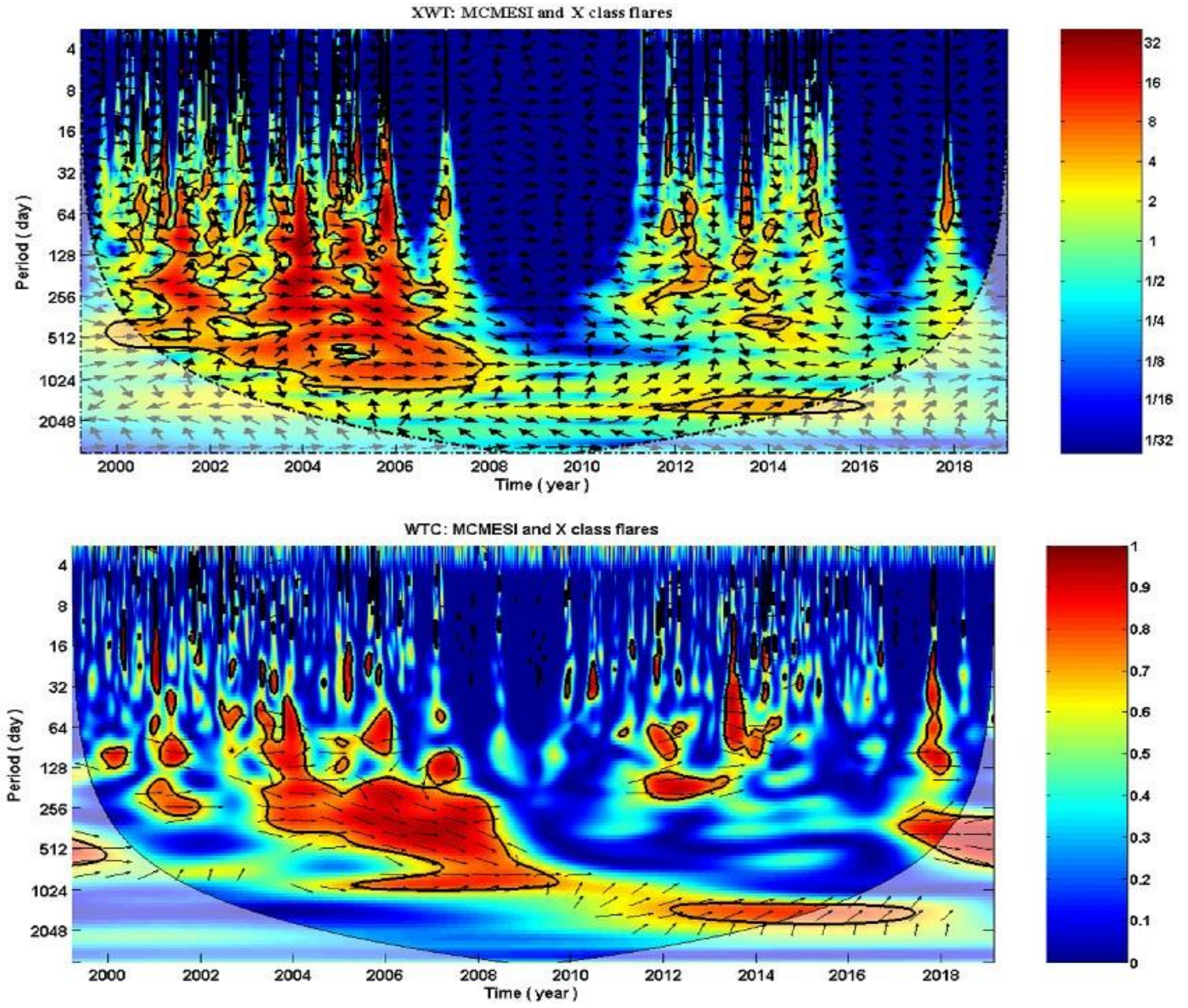

Figure 6: The same as Figure 4 but for X class flares.

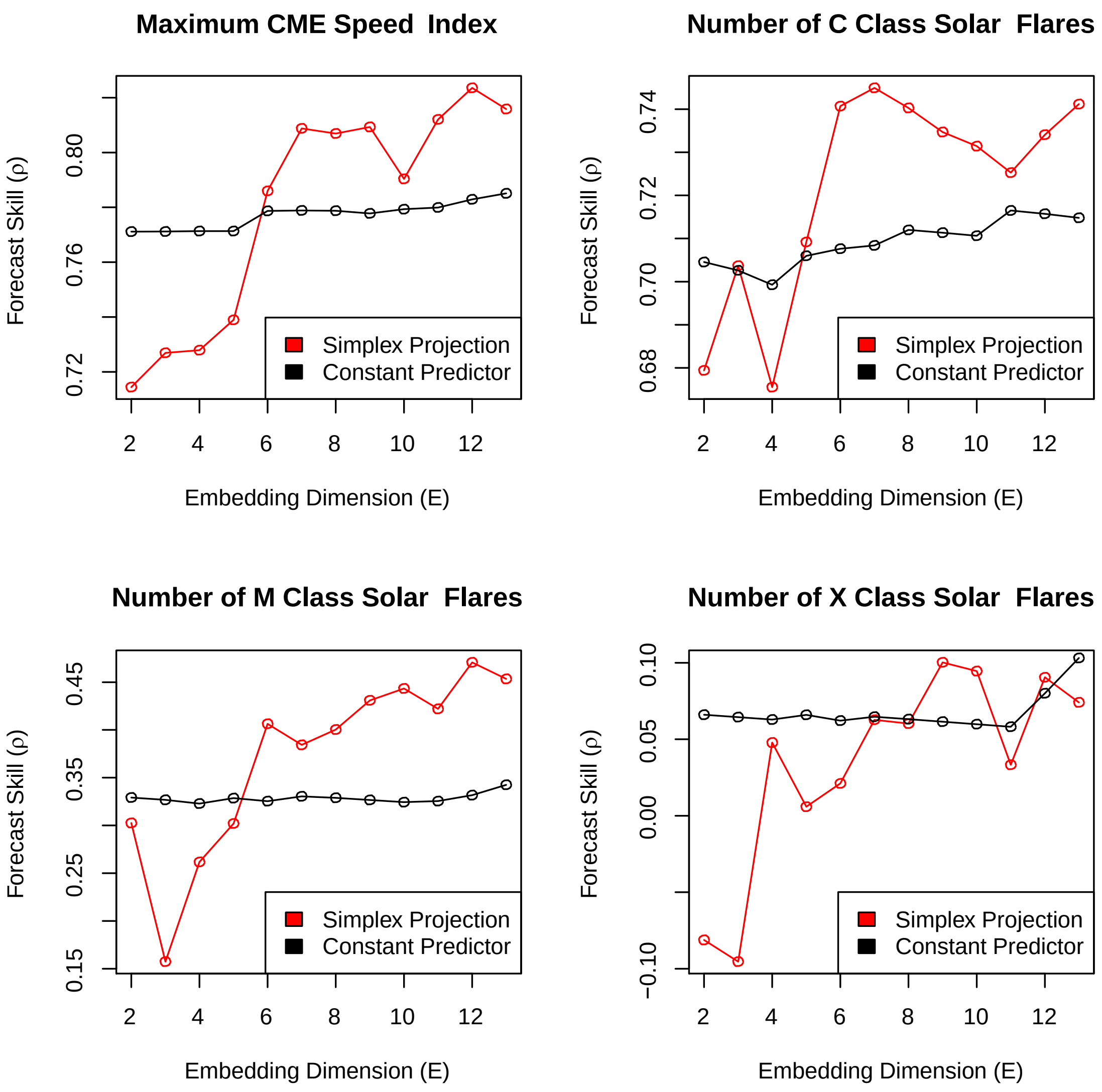

Figure 7: Correlation Coefficients between predicted and observed values of the studied datasets as a function of their embedding dimension.

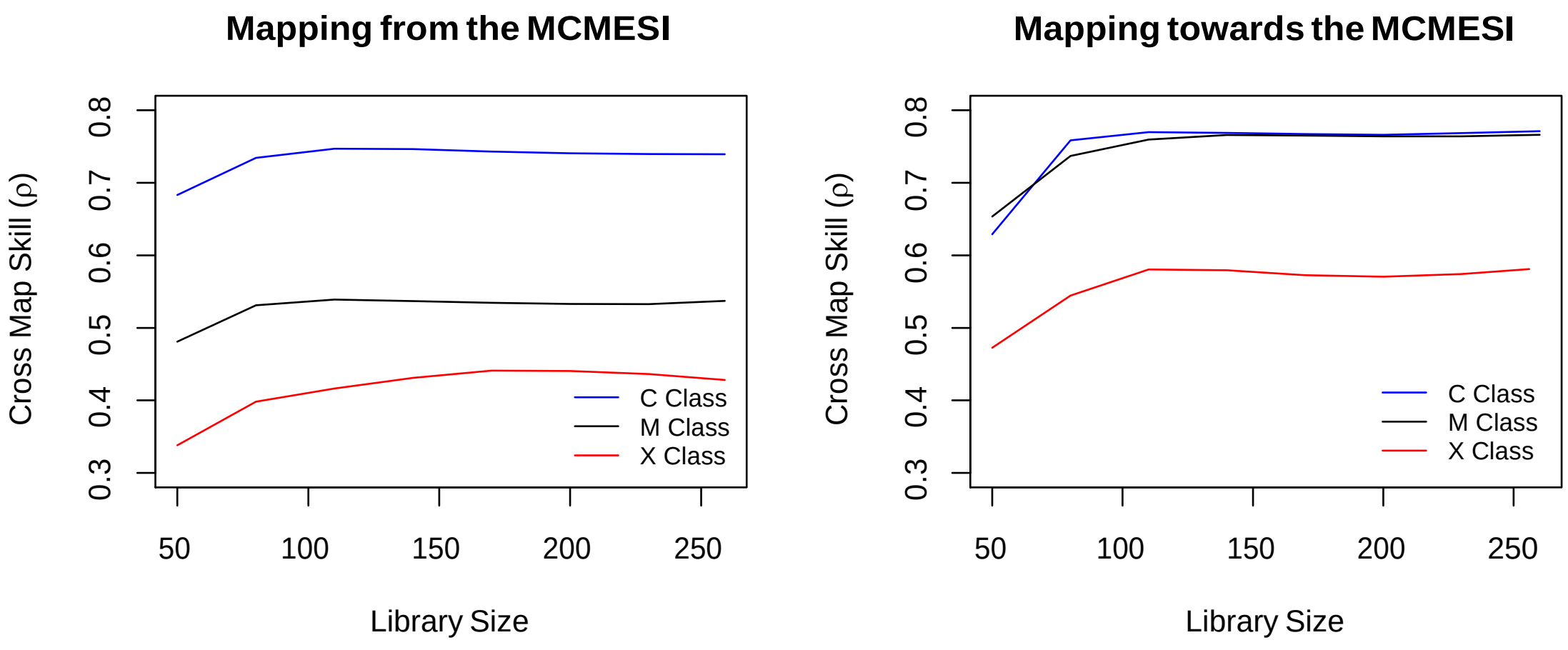

Figure 8: CCM results between different class flare numbers and the MCMESI.